\documentclass[a4paper]{jpconf}
\usepackage{graphicx}
\usepackage{amsmath}
\usepackage{cite}
\begin{document}
\title{A predictive phenomenological tool at small Bjorken-$x$}

\author{ \underline{Jos\'e Guilherme Milhano}$^{1,2}$, Javier L. Albacete$^{3}$, N\'estor Armesto$^{4}$, Paloma Quiroga-Arias$^{4}$, and Carlos A. Salgado$^{4}$}

\address{
$^1$ CENTRA, Departamento de F\'isica, Instituto Superior T\'ecnico (IST),
Av. Rovisco Pais 1, P-1049-001 Lisboa, Portugal\\
$^2$ Physics Department, Theory Unit, CERN, CH-1211 Gen\`eve 23, Switzerland\\
$^3$ Institut de Physique Th\'eorique, CEA/Saclay, 91191 Gif-sur-Yvette cedex, France. URA 2306, unit\'e de recherche associ\'ee au CNRS.\\ 
$^4$ Departamento de F\'isica de Part\'iculas and IGFAE, Universidade de Santiago de Compostela 15706 Santiago de Compostela, Spain.
}

\ead{guilherme.milhano@ist.utl.pt}

\begin{abstract}
We present the results from global fits of inclusive DIS experimental data using the Balitsky-Kovchegov equation with running coupling.

\end{abstract}


On rather general physical grounds, one expects that the large gluon density that develops in the proton (nucleus) wave function with decreasing Bjorken-$x$ will lead to deviations from standard collinear factorization. 
The necessary unitarity of the theory requires the rate of growth of the gluon density  to be limited. Such a taming arises naturally once non-linearities, due to gluon overlap, are accounted for in the evolution. 

The Colour Glass Condensate (CGC) \cite{Kovner:2000pt,Iancu:2000hn,Ferreiro:2001qy}, along with its B-JIMWLK evolution equations \cite{JalilianMarian:1997gr,Balitsky:1995ub},  provides a solid theoretical framework in which to address the non-linear dynamics driving the small-$x$ evolution of the proton (nucleus) wave function. 
The mathematical complexity of the CGC evolution equations has hindered its \textit{direct} phenomenological application. In face of such difficulties, the phenomenological use of the CGC has proceeded via dipole-models in which the general physical properties of the CGC have been, in one form or another, implemented in an effective manner.
The behaviour at small-$x$ of the parton distribution functions plays an important role in the computation of many observables in proton-proton, nucleus-nucleus, and proton-nucleus at the LHC. Given that evidence for small, yet systematic, deviations from NLO-DGLAP evolution have been identified in HERA DIS data \cite{Caola:2009iy}, it is mandatory that the best available and numerically tractable non-linear small-$x$ evolution tools replace \textit{CGC-motivated} models.

The simplest theoretically sound realization of the CGC, the Balitsky-Kovchegov (BK) equation \cite{Balitsky:1995ub,Kovchegov:1999yj}, although tractable from the numerical point of view, fails to reproduce experimental data unless rather unphysical assumptions are made. Early heuristic modifications of the BK equation \cite{Albacete:2004gw} indicated that NLO corrections would drive the evolution towards compatibility  with experimental data. The explicit computation of the full NLO-BK \cite{Balitsky:2006wa,Kovchegov:2006vj,Balitsky:2008zza} paved the way to the rigourous and phenomenologically relevant  use the CGC. Although these corrections present a rather complicated structure and are not presently amenable to an efficient numerical implementation, it was shown \cite{Albacete:2007yr} that considering only running coupling corrections to the BK equation accounts for most of the NLO effects. The ability of the rcBK equation to correctly describe experimental data was shown in \cite{Albacete:2009fh}. There, a global fit of the inclusive DIS structure function led to the public release of a parametrization of the dipole scattering cross section at small values of $x$.   
Thus, running coupling BK (rcBK), the most reliable and phenomenologically usable non-linear small-$x$ evolution tool, has become the standard choice for CGC computation of experimental observables (e.g., \cite{Albacete:2007sm,Albacete:2010bs,Albacete:2010pg}).

The  H1 and ZEUS combined data set for the DIS reduced cross section \cite{:2009wt} poses a series of new challenges and opportunities. On the one hand, the high-accuracy of the data puts the rcBK equation under much more stringent test conditions and will eventually lead to a better constrained parametrization. On the other hand, the data being given directly in terms of the reduced cross section, the experimentally measured quantity, eliminates the theoretical bias associated with the extraction of $F_2$ and $F_L$ from data. Here, we present the first results obtained both from the global fit of these data within the setup of \cite{Albacete:2009fh} where only light quarks were taken as contributing to the DIS cross section, and from a global fit in which available $F_2^c$ data is used to constrain the heavy quark contribution to the DIS cross section.

In the kinematical domain ($Q^2 < 50$ GeV, $x<0.01$) considered in this work, the reduced DIS cross-section $\sigma_r$ can be written in terms of the virtual photon-proton cross-section $\sigma_{T,L}$, with $T$ (transverse) and $L$ (longitudinal) the polarization of the virtual photon, as
\begin{equation}
\label{eq:sigred}
	 \sigma_r (x,y,Q^2) = \frac{Q^2}{4\,\pi^2\alpha_{em}}\Bigg(\sigma_{T} + \frac{2 (1-y)}{1+(1-y)^2} \sigma_{L}\Bigg)\, .
\end{equation}	 
Here, $y=Q^2/sx$ ($\sqrt{s}$  is the center of mass collision energy) is the inelasticity variable. In the dipole formulation of QCD, valid for small-$x$, one has 
\begin{equation}
  \sigma_{T,L}(x,Q^2)=\sum_f \,  \sigma_{0,f}\, \int_0^1 dz \,d{\bf r}\,\vert
  \Psi_{T,L}^f(e_f,m_f,z,Q^2,{\bf r})\vert^2\,
  {\cal N}({\bf r},x)\,,
\label{dm1}
\end{equation}
where $\Psi_{T,L}^f$ is the light-cone wave function for a virtual photon to fluctuate into a quark-antiquark dipole of quark flavor $f$ (with mass $m_f$ and electric charge $e_f$). ${\cal N}({\bf r},x)$ is the imaginary part of the dipole-target scattering amplitude averaged over impact parameter, with $\bf r$ the transverse dipole size. $\sigma_{0,f}$ is (half) the transverse area over which quarks of a given flavour are distributed. Light quarks are taken to be identically distributed ($\sigma_{0,f=u,d,s} =\sigma_{0,light} $). When accounting for heavy flavour contributions in (\ref{eq:sigred}) we allow $\sigma_{0,f=c,b} =\sigma_{0,heavy}$ to be different from $\sigma_{0,light}$. These are free fit parameters. 

The evolution of  ${\cal N}({\bf r},x)$ is given by the rcBK equation:
\begin{multline}
  \frac{\partial{\cal{N}}(r,x)}{\partial\,\ln(x_0/x)}
  =  \frac{N_c\,\alpha_s(r^2)}{2\pi^2}\, \int d{\bf r_1}\,
  \left[\frac{r^2}{r_1^2\,r_2^2}+
    \frac{1}{r_1^2}\left(\frac{\alpha_s(r_1^2)}{\alpha_s(r_2^2)}-1\right)+
    \frac{1}{r_2^2}\left(\frac{\alpha_s(r_2^2)}{\alpha_s(r_1^2)}-1\right)
  \right]  \\
 \cdot
 \left[{\cal N}(r_1,x)+{\cal N}(r_2,x)-{\cal N}(r,x)-
    {\cal N}(r_1,x)\,{\cal N}(r_2,x)\right]\,.
\label{bkrun}
\end{multline}
where ${\bf r_2}={\bf r}-{\bf r_1}$, and $x_0$  (in our case $x_0=0.01$) is the value of $x$ where the evolution starts (the highest value of $x$ in the data included in the fits). 
The kinematic cost of the virtual photon to fluctuate into a dipole of a given mass is accounted for by $x=x_{exp}\,(1+ 4\,m_f^2/Q^2)$. The running coupling in (\ref{bkrun}) is evaluated at 1-loop accuracy in coordinate space
\begin{equation}
\label{eq:rc}
	\alpha_{s,n_f} (r^2) =  \frac{4\pi}{\beta_{0, n_f} \ln \Big(\frac{4 C^2}{r^2 \Lambda^2_{n_f}}\Big) }\, ,
	\qquad \beta_{0,n_f} = 11 - \frac{2}{3} n_f \, ,
\end{equation}
where the constant $C$ (a fit parameter) accounts for the uncertainty in the Fourier transform from momentum to coordinate space. 
When only light flavours ($f=u,d,s$) are taken as contributing to (\ref{eq:sigred}),  the coupling (\ref{eq:rc}) is evaluated with $n_f = 3$ (\textit{light only}). Once the contribution of heavy flavours   ($f=c,b$) is accounted for, the coupling (\ref{eq:rc}) is evaluated in a variable flavour scheme matched on the quark mass thresholds (\textit{light+heavy}). Since in the rcBK equation  (\ref{bkrun}) all dipoles sizes are explored, an infrared regulation is called for: $\alpha_s (r^2 > r^2_{fr}) = \alpha_{fr}=0.7$ where $r^2_{fr}$ is the dipole size at which the coupling reaches $\alpha_{fr}$. Although we have explored \cite{best} a number of different initial conditions, the results show a negligible dependence on specific choice. For brevity, only results obtained with the GBW initial condition
\begin{equation}
	 {\cal N^{GBW}}(r,x\!=\!x_0)=
1-\exp{\left[-\frac{\left(r^2\,Q_{s\,0}^2\right)^{\gamma\,}}{4}\right]}\, ,
\end{equation}
are shown here. The fit parameters $Q_{s\,0}^2$ and $\gamma$ are, respectively, the saturation scale and the characteristic fall-off of the dipole scattering amplitude with decreasing $r$  at the initial $x_0$.

\begin{figure}[h]
\begin{minipage}{10cm}
\includegraphics[width=10cm]{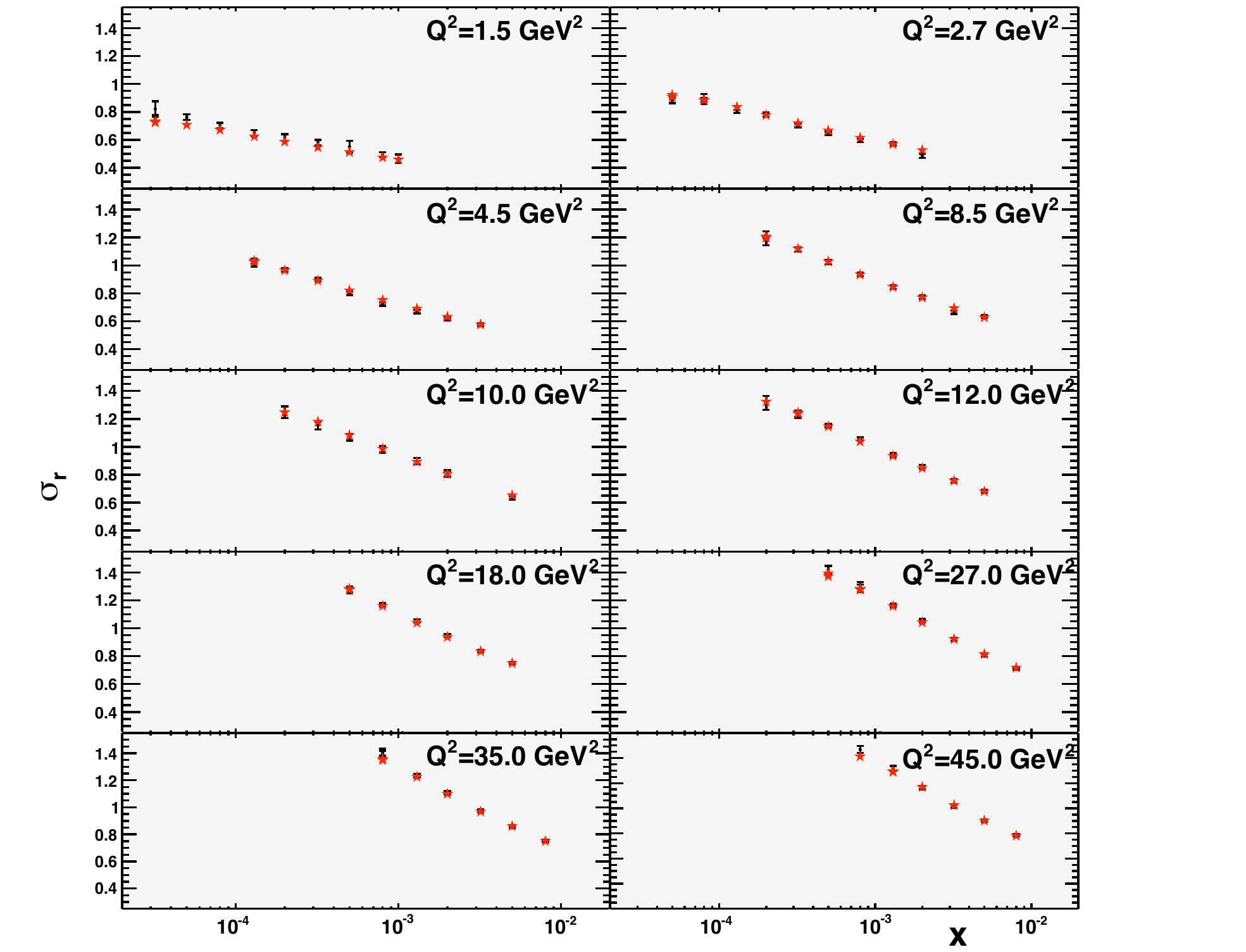}

\end{minipage}
\begin{minipage}{5cm}
\vspace{3cm}
\caption{\label{Fig1} Comparison of reduced cross section experimental data (in black)  with our (\textit{light only}) fit results (red stars). Notice that the experimental error is often smaller that size of the symbols used for the data points.}
\end{minipage}
\end{figure}

Fig. \ref{Fig1} shows a comparison of data for the reduced cross section (selected $Q^2$  data bins) with the results of a \textit{light only} fit using all combined H1/ZEUS data (196 points) within cuts $Q^2 <50$ GeV$^2$ and  $x<0.01$.
The mass of the light flavours was kept fixed,  $m_{light}=0.14$ GeV. The obtained fit parameters ---  $\sigma_{0,light}=33.1$ mb, $Q_{s\,0, light}^2 = 0.26$ GeV$^2$, $\gamma_{light}=0.97$, and  $C = 2.3$ --- give $\chi^2/{\rm d.o.f.} =206/192=1.07$. 

\begin{figure}[h]
\begin{minipage}{10cm}
\includegraphics[width=10cm]{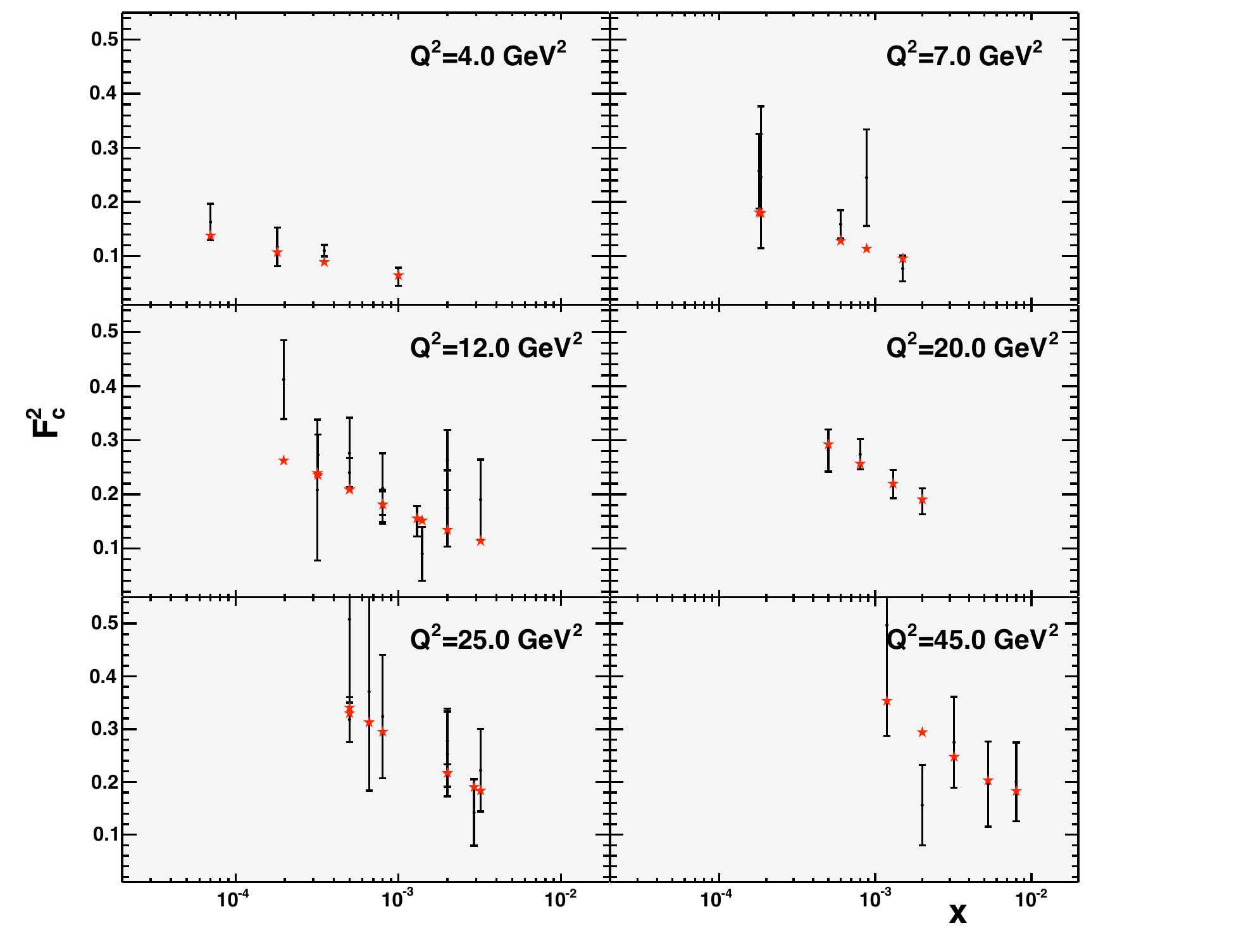}
\end{minipage}
\begin{minipage}{5cm}
\vspace{3cm}
\caption{\label{Fig2} Comparison of  $F_{2}^c$ data (in black)  with our (\textit{light+heavy}) fit results (red stars). \textit{light+heavy} fits give description of reduced cross section similar to that shown in  Fig. \ref{Fig1} }
\end{minipage} 
\end{figure}

Fig.  \ref{Fig2} shows a comparison of data on $F_{2}^c$ (the charm contribution to $F_2$) with the results of a \textit{light+heavy} fit. In this case, all existing data on  $F_{2}^c$ (within the same cuts as above) is added to the H1/ZEUS sample (a total of 265 points). We allow for independent light and heavy initial conditions and  for a free light flavour mass. The heavy quark masses are set to $m_c=1.4$ GeV and $m_b=4.75$ GeV.
The $\chi^2/{\rm d.o.f.} =312/257=1.2$ is obtained for the fit parameters $\sigma_{0,light}=32.7$ mb,  $Q_{s\,0, light}^2=0.26$ GeV$^2$, $\gamma_{light}=1.24$, $\sigma_{0,heavy}= 28.4$ mb, $Q_{s\,0, heavy}^2= 0.18$ GeV$^2$, $\gamma_{heavy}= 0.95$, $C=3.3$, and $ m_{light} = 0.104$ GeV.

The ability of the rcBK equation to accurately describe the high precision H1/ZEUS combined sample and the charm component of the structure function strengthens its role as a viable phenomenological tool to address the small-$x$ behaviour of the proton structure. Further details, along with a parametrization of the dipole-proton cross section for use in the computation of physical observables, will be given in \cite{best}.

\ack
We acknowledge support from MICINN (Spain) under project FPA2008-01177 and FPU grant; Xunta de Galicia (Conselleria de Educacion) and through grant PGIDIT07PXIB206126PR, the Spanish Consolider-
 Ingenio 2010 Programme CPAN (CSD2007-00042) and Marie Curie MEST-CT-2005-020238-EUROTHEPHY (PQA); and Funda\c c\~ao para a Ci\^encia e a Tecnologia (Portugal) under project CERN/FP/109356/2009 (JGM).

\section*{References}
\bibliography{bib}

\end{document}